\documentstyle[11pt,newpasp,twoside,epsf]{article}
\def\edcomment#1{\iffalse\marginpar{\raggedright\sl#1\/}\else\relax\fi}
\reversemarginpar

\begin{document}
\title{Hubble Space Telescope Imaging of the Young Planetary Nebula GL 618}
\author{Susan R. Trammell}
\affil{Physics Department, University of North Carolina at Charlotte, 9201 University City Blvd., Charlotte, NC 28223}

\begin{abstract}
We present narrow-band Hubble Space Telescope images of the
young planetary nebula GL 618. These images have allowed us to
study the detailed morphology of shock-excited emission present in
the bipolar lobes of this object. These images reveal the presence of
three highly collimated outflows emanating from the central regions
of GL 618. We discuss the significance of the detection of these
outflows and the possible origins of these features.
\end{abstract}

\section{Introduction}

GL 618 is a young, bipolar planetary nebula (PN). Ground-based
optical and near-IR imaging of this object reveal two lobes of
emission (each about 3$\arcsec$ in extent) separated by a dark lane.
The central regions of the nebula are hidden from direct view at optical
wavelengths by the lane of obscuring material. 
The spectrum of GL 618 is composed of a faint continuum
and a variety of low-excitation emission lines. Trammell, Dinerstein,
\& Goodrich (1993) used spectropolarimetry to study GL 618 and
found that the continuum and part of the permitted line emission are
reflected from deep in the nebula.
The low-excitation, forbidden line flux and remainder of the
permitted line emission are produced in the bipolar lobes. The
emission produced in the bipolar lobes is indicative of shock heating
($V_s$ = 50$-$100 kms$^{-1}$). Long-slit optical spectroscopy of GL 618
confirms that the shock emission is associated with out-flowing gas
(Carsenty \& Solf 1982) and near-IR spectroscopy of GL 618 has
revealed the presence of thermally excited H$_2$ emission
(Thronson 1981; Latter et al. 1992). GL 618 exhibits [Fe II]
emission also thought to be associated with the shock-heated gas
(Kelly, Latter, \& Rieke 1992) and more recent observations
establish that this emission is associated with an outflow (Kelly,
Hora, \& Latter 1999). Shock-excited emission dominates the
spectra of the lobes of GL 618. We present WFPC2 images of this
object that demonstrate that the source of this shock-excited emission is a
set of highly collimated outflows originating in the central regions of 
the object.

\section{Observations}

We have obtained WFPC2 images of GL 618 as part of a HST
Cycle 6 program. GL 618 was centered in the Planetary Camera
which has a 36$\arcsec$ $x$ 36$\arcsec$ field of view and a plate scale
of 0.0455\arcsec per pixel$^{-1}$. Images were obtained through four
filters: F631N (isolating [O I]$\lambda$6300 line emission), F656N
(isolating H$\alpha$ line emission), F673N (isolating [S II]
$\lambda$$\lambda$6717,31 line emission), and F547M (a continuum band). These
filters were chosen so that we could study the morphology of the
shock-excited emission in the lobes of GL 618. The images were
processed through the HST data reduction pipeline procedures and
cosmic rays were removed by combining several exposures of each
object. The images of GL 618 were obtained on 23 October 1998
and exposure times ranged from 15 to 45 minutes.

\section{Results}

The overall morphologies seen in the [S II] and [O I] images are
similar (Figure 1, panels (a) and (b)). These images trace the
morphology of the shock-excited forbidden line emission in GL
618. Three highly collimated outflows, or jets, are seen in both
images. The brightest emission occurs near the tip of each of the
outflows and there is no forbidden line emission seen in the central
regions of GL 618. Ripple-like morphology is seen in the outflows
in both the [S II]  and [O I]  images. These ripples might be the
result of instabilities in the flow and/or an interaction with the
surrounding nebular material.  

The morphology observed in the H$\alpha$ image (Figure 1, panel (c))
differs slightly
from the forbidden line morphology.
H$\alpha$ emission is seen associated with the outflows, but in
addition, a significant amount of H$\alpha$ emission is seen towards
the central regions of GL 618. Spectropolarimetric observations
indicate that part of the H$\alpha$ emission is reflected and part of
this emission is produced by shocks in the lobes (Trammell et al.
1993). We have spatially separated these components in the HST
images. The H$\alpha$ emission associated with the central regions
of the object is probably reflected emission from an H II region buried
deep in the nebula. A high density H II region has been observed at
the center of GL 618 at radio wavelengths (e.g. Kwok \& Feldman
1981) and in the reflected optical spectrum (Trammell et al. 1993).
The H$\alpha$ emission coincident with the outflows is the
shock-excited component of the permitted line emission.

The [O I] to [S II] line ratios
in the bullet-like structures at the tips of the outflows
are approximately 3.0$-$3.5. By
comparing these observed line ratios with the predictions of planar
shock models (Hartigan, Raymond, \& Hartmann 1987), we
estimate the shock velocity in these regions to be approximately 80
kms$^{-1}$.  This is consistent with the range in shock velocities
estimated from previous spectropolarimetric observations (Trammell et al.
1993).

Careful examination of the bullet-like structures at the tip of the
outflow in the upper lobe in Figure 1 reveals an excitation gradient
across this region. H$\alpha$ is brightest on the side of the spot
facing away from the central regions of GL 618. [S II] and [O I]
are brighter on the side closest to the central source. This type of
gradient is expected for a jet flowing away from the central source
and impinging on the surrounding nebular material. The bright spots
near the tops of the outflows are not clumps of material being
overrun by a wind or outflow.

\section{Discussion}

HST
observations (e.g. Trammell \& Goodrich 1996; Sahai \& Trauger
1998) and ground-based imaging surveys (e.g. Balick 1987;
Schwarz, Corradi, \& Melnick 1992) have revealed the presence of
collimated outflows, FLIERs, and a myriad of other small-scale
structures in PN. The origins of these structures and their role in the
overall development of PN remain puzzling. 
The debate concerning the origin of these small-scale structures, and
also the 
formation of aspherical PN in general,
centers on whether binary or
single stars are responsible for producing aspherical mass loss.
Both models of binary
star interaction (e.g. Soker \& Livio 1994) and magnetic
confinement (e.g. Garcia-Sergura 1997), while providing a scheme
for producing the overall aspherical structure in PN, may also
provide mechanisms to produce the highly collimated outflows.
The complex, mulitpolar outflow geometry seen in GL 618 may be difficult
for either of these types of models to explain.

Our observations demonstrate that jets can be present during the early
phases of PN development and may play an important role in the
early shaping of these objects. Futher, these collimated outflows may set
the stage for the development of other small scale structures seen in
more evolved objects.

\vspace{0.2in}
\noindent
Support for this research was provided by NASA through
grant number GO-06761 from the
Space Telescope Science Institute, which is operated by AURA, Inc., under
NASA contract NAS 5-26555.
\newpage

\begin{figure}
\plotone{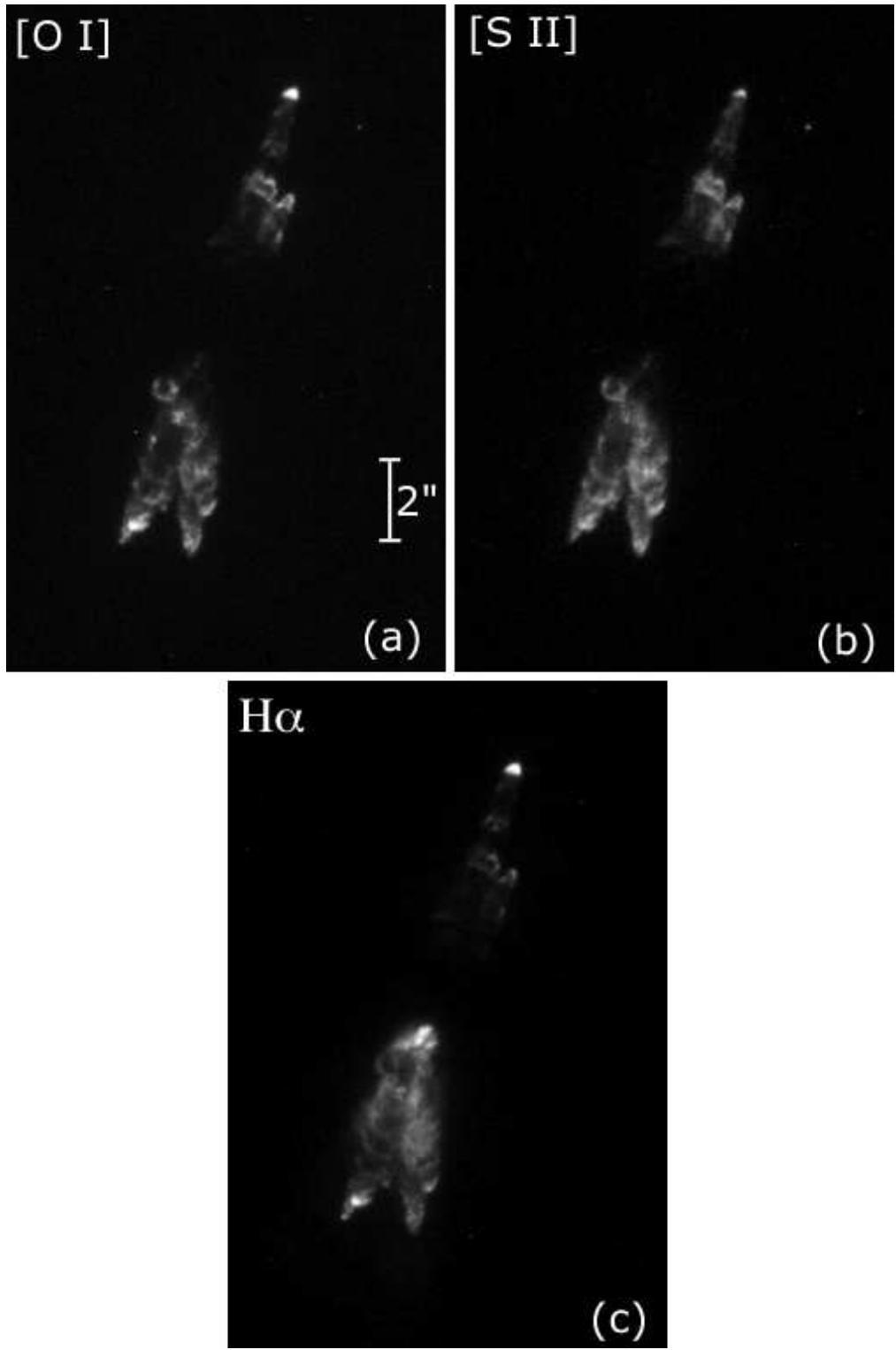}
\caption{Narrow-band WFPC2 images of GL 618. Three collimated outflows are clearly evident in all three images. (a) The F631N image which traces the [O I] emission (b) The F673N filter image which traces the [S II] emission. (c) The F656N filter image which traces the H$\alpha$ emission.}
\end{figure}

\end{document}